\documentclass[journal]{IEEEtran}
\usepackage{cite}
\usepackage{amsmath,amssymb,amsfonts}
\usepackage{algorithmic}
\usepackage{graphicx}
\usepackage{textcomp}
\usepackage{xcolor}
\usepackage{booktabs} 
\usepackage[ruled]{algorithm2e}
\usepackage{hyperref}
\usepackage{stfloats}
\usepackage{makecell}
\usepackage{amsthm}
\usepackage[caption=false,font=normalsize,labelfont=sf,textfont=sf]{subfig}

\allowdisplaybreaks[4]

\begin{document}

\title{Large Language Model Assisted  Optimal Bidding of BESS in FCAS Market: An AI-agent based Approach}

\author{Borui~Zhang,~\IEEEmembership{{Graduate Student Member},~IEEE,}
        Chaojie~Li,~\IEEEmembership{Member,~IEEE,} 
        Guo~Chen,~\IEEEmembership{Member,~IEEE,}       and~Zhaoyang~Dong,~\IEEEmembership{Fellow,~IEEE}%Zhao~Xu,~\IEEEmembership{Senior Member,~IEEE,} and~Zhaoyang~Dong,~\IEEEmembership{Fellow,~IEEE}% <-this % stops a space
\thanks{B. Zhang, C. Li and G. Chen are with the School of Electrical Engineering and Telecommunications, UNSW Sydney, NSW 2052, Australia. (borui.zhang;chaojie.li;guo.chen@unsw.edu.au)}
%\thanks{Z. Xu is with the Department of Electrical Engineering, The Hong Kong Polytechnic University, Hung Hom, Kowloon, Hong Kong. (zhao.xu@polyu.edu.hk)}
\thanks{Z. Dong is with the Department of Electrical Engineering,  City University of Hong Kong, Hong Kong. (zy.dong@ieee.org) }
}

\markboth{Journal of \LaTeX\ Class Files,~Vol.~14, No.~8, August~2015}%
{Shell \MakeLowercase{\textit{et al.}}: Bare Demo of IEEEtran.cls for IEEE Journals}

\maketitle

\begin{abstract}
To incentivize flexible resources such as Battery Energy Storage Systems (BESSs) to offer Frequency Control Ancillary Services (FCAS), Australia's National Electricity Market (NEM) has implemented changes in recent years towards shorter-term bidding rules and faster service requirements. However, firstly, existing bidding optimization methods often overlook or oversimplify the key aspects of FCAS market procedures, resulting in an inaccurate depiction of the market bidding process. Thus, the BESS bidding problem is modeled based on the actual bidding records and the latest market specifications and then formulated as a deep reinforcement learning (DRL) problem. Secondly, the erratic decisions of the DRL agent caused by imperfectly predicted market information increases the risk of profit loss. Hence, a Conditional Value at Risk (CVaR)-based DRL algorithm is developed to enhance the risk resilience of bidding strategies. Thirdly, well-trained DRL models still face performance decline in uncommon scenarios during online operations. Therefore, a Large Language Models (LLMs)-assisted artificial intelligence (AI)-agent interactive decision-making framework is proposed to improve the strategy timeliness, reliability and interpretability in uncertain new scenarios, where conditional hybrid decision and self-reflection mechanisms are designed to address LLMs' hallucination challenge. The experiment results demonstrate that our proposed framework has higher bidding profitability compared to the baseline methods by effectively mitigating the profit loss caused by various uncertainties.

\end{abstract}


\vspace{-3mm}
\section{Introduction}
\IEEEPARstart{I}{n} the pursuit of mitigating climate change and transitioning to low-carbon, sustainable goals, renewable energy development has advanced yet its extensive integration into power grids has significantly impacted system stability and security. In Australia's National Electricity Market (NEM), the Frequency Control Ancillary Services (FCAS) market purchases frequency regulation (FR) capacities from bidders to ensure stable system frequency \cite{RN123}. In 2023, a new 1-second FCAS market \cite{RN124} was introduced in NEM to incentivize flexible resources such as Battery Energy Storage Systems (BESSs) to offer responses within one second. Considering the high cost of BESSs \cite{6589173}, it is imperative to develop highly profitable bidding strategies.

%Adopting an aggressive bidding strategy can lead to higher profits but with increased risks, whereas a conservative approach offers more stable but potentially lower income. 
%Additionally, in practical applications, bidders often participate in multiple markets, such as joint arbitrage and FR markets \cite{PUSCEDDU2021116274}. The coupling restrictions and the aging cost of BESS makes the optimization problem of bidding strategies more complex, necessitating more intelligent and efficient solutions.

Existing studies have focused on the optimization of bidding strategies in ancillary service markets. The study in \cite{sgmkt} adopts a ``predict-then-optimize" framework for a single system in a single market, utilizing linear programming (LP) based on predicted prices to seek the optimal bidding strategy of BESS in the energy market. Furthermore, \cite{FENG2022105508} applies a genetic algorithm (GA) to address the bidding problem of a wind-BESS system in multiple markets, considering the joint bidding capacity allocation in energy and frequency reserve markets. \cite{WANG2023121918} models a stochastic game to find Nash equilibrium bidding solutions for multiple virtual power plants (VPPs) in energy and frequency markets. However, mathematical methods are constrained by low efficiency and an over-reliance on forecasting, hence facing challenges in making timely bidding decisions with high returns in dynamic market conditions \cite{GUO2022121873}. More importantly, existing solutions are not applicable in the real-world FCAS market because the actual bidding procedure has various compulsory requirements, such as bids must be step-like over multiple bands, and bidding price modification is prohibited in the real-time market which is often ignored or oversimplified. Bidders are also often assumed to be price-takers, omitting the responses of market operators.

Artificial intelligence (AI)-based methods, such as reinforcement learning (RL) and deep RL (DRL), have been applied to bidding optimization in ancillary markets \cite{9483690,9803252} and can improve efficiency and get rid of the over-reliance on forecasting by learning from historical experiences \cite{9016168}. In \cite{9712169}, inverse RL (IRL) is used to seek the objective of BESSs based on bidding data in energy and ancillary service markets. Additionally, \cite{liu2022optimal} utilizes multi-agent RL (MARL) to address the bidding game among multiple VPPs while ensuring fairness among bidders.
%These studies have laid a foundational framework and solution methods for bidding strategy optimization in the ancillary markets. 
However, the aleatoric uncertainty arising from dynamic environments and the epistemic uncertainty caused by limited training data remain challenges limiting the practical applications of DRL \cite{clements2019estimating}. 

In DRL bidding problems, the aleatoric uncertainty comes from uncertain market behaviors and imperfect insight and results in DRL agents making erratic decisions under imperfectly predictable information. Conditional value-at-risk (CVaR), as a popular risk measure, can enhance economic benefits by managing market risk \cite{ALEXANDER2006583}. In \cite{WANG2023121918}, CVaR is integrated into a risk-averse mathematical model to reduce the risk of overbidding. If CVaR is incorporated into the DRL model to control the risk of erratic decisions of the DRL agent, the profit loss due to the aleatoric uncertainty can be effectively mitigated.

 %Unfortunately, the FCAS market conditions change significantly over time, exposing the weakness of DRL-trained knowledge in terms of timeliness and causing a decline in profitability. Even though the model can be corrected online by penalizing inferior bidding strategies, the profit loss has already occurred.

The epistemic uncertainty is from uncommon market conditions outside the training set and exposes the weakness of DRL agents performing well during training but poorly in new environments \cite{lockwood2022review}. \cite{cao2024survey} suggests that Large Language Models (LLMs) are promising in driving DRL agents to make more reliable decisions in new environments, leveraging their rich knowledge and natural language processing capabilities. \cite{sharan2023llm} proposes an LLMs-assist autonomous driving planning to better handle uncertain driving scenarios. \cite{nascimento2023gpt} suggests a ``GPT-in-the-loop" method, utilizing GPT to aid agent systems in interactive decision-making for better performance and interpretability. Thus, if unfaithful decisions by LLMs' hallucination \cite{ji-etal-2023-towards} can be mitigated, it will be a great opportunity to reduce DRL's profit loss in unfamiliar market conditions through interactive collaboration between DRL and LLMs.

Therefore, this paper establishes models that comply with the actual rules of NEM, proposes a CVaR-DRL algorithm to seek the optimal bidding strategies to improve the profitability of BESS under market uncertainty and dynamic bidding risks, and builds an LLMs-assisted AI-agent interactive decision-making framework to enhance the bidding performance of the DRL agent in new scenarios through LLMs-assisted AI-agent interactive collaboration. Fig.~\ref{fig:1} shows the framework of the bidding system. The main contributions are as follows:

\begin{figure*}[htbp]
\centerline{\includegraphics[width=0.85\linewidth]{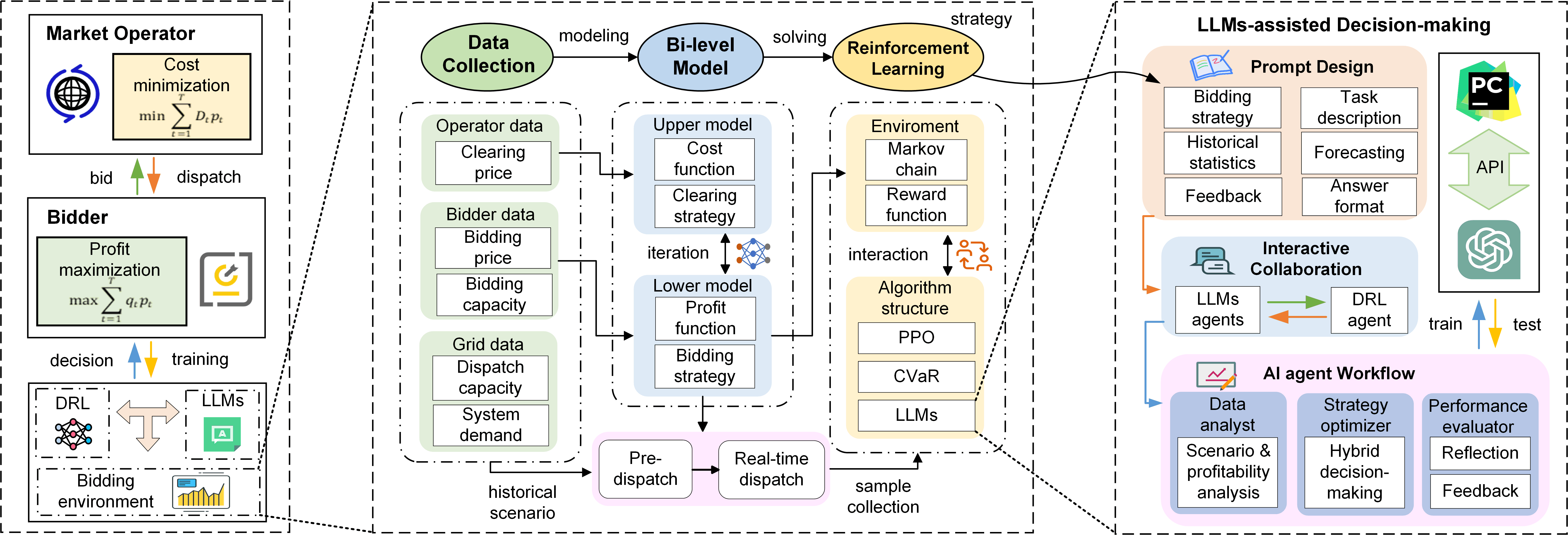}}
\caption{LLMs-based BESS Ancillary Service Autonomous Intelligent Bidding System.}
\label{fig:1}
\end{figure*}

\begin{itemize}
\item To address the inability of existing bidding solutions based on models with oversimplified market procedures, price taker assumptions, linear approximations, etc., to cope with practical requirements, a bi-level bidding game problem between the market operator and the price-maker BESS is modeled based on the actual bidding records of each bidder in the FCAS market and the NEM's two-stage multi-band bidding requirements.

\item To handle the weakness of traditional mathematical and DRL methods in terms of efficiency, performance and risk resilience, a CVaR-DRL model is proposed for bidding optimization to manage the risk of erratic decisions of the DRL agent and reduce the profit loss caused by uncertain market conditions and imperfect insight.

\item To improve the strategy timeliness, reliability, and interpretability of DRL in unfamiliar scenarios, an LLMs-assisted AI-agent interactive decision-making framework is built for online market analysis, assisted decision and interpretable feedback, where conditional hybrid decision and self-reflection mechanisms are designed to mitigate unfaithful decisions by LLMs' hallucination.

\end{itemize}

%The remainder of the paper is organized as follows. Section II introduces the preliminaries of our research. Section III presents the DRL problem formulation and our proposed algorithm. In Section IV, the results of the case study are analyzed. Section V concludes the paper.
The paper is organized as follows: Section II covers research preliminaries. Section III presents the DRL problem and proposed algorithm. Section IV analyzes case study results. Section V concludes the paper.

\section{Preliminaries}
%In this section, the background, definitions, variables and models for the optimal bidding problem formulation are introduced.
% \vspace{-3mm}

\subsection{FCAS Markets Rules and Mechanisms}
In Australia's NEM, the Australian Energy Market Operator (AEMO) ensures that the system frequency remains within the normal operation band by purchasing sufficient capacity from FCAS providers' bids to match the FCAS demand \cite{RN121}. FCAS is divided into regulation and contingency services: regulation FCAS is dispatched to correct minor frequency deviations while contingency FCAS restores larger frequency deviations when the frequency exceeds the standard band \cite{RN123}. Depending on the directions of frequency correction, they can be further classified as raised services and lower services. The specific market operation has the following stages.

\subsubsection{Bidding Stage} The NEM operates a real-time spot market only but has a pre-dispatch stage and a real-time stage \cite{RIESZ201586}. In the pre-dispatch stage, bidders are required to submit the next 24-hour bids one day in advance. A standard bid has 10 bidding bands including 10 bidding capacities and prices, and they must be monotonically increasing. In the real-time stage, the bidding prices will no longer be modified but the capacities can be flexibly adjusted via re-bids before the next trading interval \cite{RN123}. This rule allows bidders to adjust the capacity allocation in dynamic real-time market conditions while ensuring the clearing prices do not experience undesired large fluctuations due to re-bidding.

\subsubsection{Clearing Stage} In the market clearing stage, AEMO aims to purchase sufficient bidding capacities with minimal costs. In each trading interval, AEMO will sort all bids by price and prioritize enabling bids with lower prices until the dispatched capacity exactly matches the FCAS demand. Subsequently, the highest price among all enabled bids will be selected as the clearing price \cite{RN123}. It is worth noting that AEMO considers each bidder's capacity constraints across the energy, regulation, and contingency markets, meaning that the total dispatched capacity of this bidder across these three markets cannot exceed its maximum output power limit \cite{RN122}.

\subsubsection{Settlement Stage} In the settlement stage, enabled bidders will be settled based on the enabled capacity and the clearing price. Note that the enabled capacity is different from the actually dispatched capacity. Even if a bidder is on standby as reserve capacity, it will obtain the same revenue \cite{RN123}.

\subsection{Market Clearing Model}
At each trading interval $t$, AEMO will deploy sufficient capacity from $N$ bids to meet the demand $d_t$ in the order of the cost. $bc_{t,n,k}$ and $bp_{t,n,k}$ denote the $k$-th bidding capacity and bidding price respectively of the $n$-th bidder at time $t$. Then AEMO will determine the enabled capacity $ec_{t,n}$ and the clearing price $cp_t$ which will be settled to the $n$-th bidder. Let $m\in \{lr,rr,lc,rc\}$ denote the lower regulation, raise regulation, lower contingency, and raise contingency FCAS markets respectively. The market clearing model is as follows:

\begin{subequations}
\begin{small}
\begin{alignat}{2}
\begin{split}
&\min\quad\sum_{m}\sum_{t=0}^{T}(cp_t^m d_t^m)
\end{split}  \label{eq:obj1}\\
\mbox{s.t.}\quad
&\sum_{k=1}^{10}u_{t,n,k}^m\le1, &\label{1b}\\
&BC_{t,n}^m=\sum_{k=1}^{10}u_{t,n,k}^m bc_{t,n,k}^m,&\label{1c}\\
&0 \le ec_{t,n}^m\le BC_{t,n}^m,&\label{1d}\\
&BP_{t,n}^m=\sum_{k=1}^{10}u_{t,n,k}^m bp_{t,n,k}^m,&\label{1e}\\
&cp_t^m\ge \sum_{k=1}^{10}u_{t,n,k}^m BP_{t,n}^m, &\label{1f}\\
&cp_t^m\le \sum_{k=1}^{10}u_{t,n,k}^m BP_{t,n}^m+M(1-v_{t,n}^m), &\label{1g}\\
&\sum_{n=1}^{N}v_{t,n}^m\ge1, &\label{1h}\\
&\sum_{n=1}^{N}ec_{t,n}^m+s_t^m=d_t^m, &\label{1i}\\
&ec_{t,n}^{lr}+ec_{t,n}^{lc}+Pc_{t,n}\le Pc_{t,n}^{max},& \label{1j}\\
&ec_{t,n}^{rr}+ec_{t,n}^{rc}+Pd_{t,n}\le Pd_{t,n}^{max}, &\label{1k}\\
&u_{t,n,k}^m,v_{t,n}^m\in \{0,1\}, &\label{1l}\\
&\forall m\in \{lr,rr,lc,rc\}, &\label{1m}
\end{alignat}
\end{small}
\end{subequations}
where the optimization objective \eqref{eq:obj1} is to minimize the costs across all FCAS markets. \eqref{1b}-\eqref{1c} aim to determine the bidding capacity $BC_{t,n}^m$ of each bidder by selecting at most one bidding band from ten bands. 
\eqref{1d} means that AEMO can flexibly enable the capacity $ec_{t,n}^m$ from the selected bidding capacity $BC_{t,n}^m$. Similar to \eqref{1b}, \eqref{1e} determines the corresponding bidding price $BP_{t,n}^m$ at the selected band. \eqref{1f}-\eqref{1h} aims to use the big M method to determine the maximum enabled bidding price (i.e. the clearing price $cp_t^m$), where M is a very large value. \eqref{1i} ensures the overall enabled FCAS capacity plus the overall market supply variation $s_t^m$ is equal to the FCAS demand $d_t^m$. \eqref{1j} ensures the joint enabled capacity of each bidder in regulation lower, contingency lower FCAS markets plus the charging capacity in energy market $Pc_{t,n}$ never exceed its maximum charging capacity $Pc_{t,n}^{max}$. Similarly, \eqref{1k} ensures the joint enabled capacity of each bidder in regulation raise, contingency raise FCAS markets plus the discharging capacity in energy market $Pd_{t,n}$ never exceed its maximum discharging capacity $Pd_{t,n}^{max}$.
As shown in \eqref{1l}, $u_{t,n,k}$ and $v_{t,n}$ are binaries.

% \begin{subequations}
% \begin{alignat}{2}
% \begin{split}
% &\min\quad\sum_{t=0}^{T}(cp_t^{lr} d_t^{lr}+cp_t^{rr} d_t^{rr}+cp_t^{lc} d_t^{lc}+cp_t^{rc} d_t^{rc})
% \end{split}  \label{eq:obj2}\\
% \mbox{s.t.}\quad
% &ec_{t,n}^{lr}+ec_{t,n}^{lc}+Pc_{t,n}\le Pc_{t,n}^{max},& \label{2b}\\
% &ec_{t,n}^{rr}+ec_{t,n}^{rc}-Pd_{t,n}\le Pd_{t,n}^{max}, &\label{2c}
% \end{alignat}
% \end{subequations}
% where the optimization objective \eqref{eq:obj2} is to minimize the costs across all FCAS markets. \eqref{2b} ensures the joint enabled capacity of each bidder in regulation lower, contingency lower FCAS markets plus the charging capacity in energy market $Pc_{t,n}$ never exceed its maximum charging capacity $Pc_{t,n}^{max}$. Similarly, \eqref{2c} ensures the joint enabled capacity of each bidder in regulation raise, contingency raise FCAS markets plus the discharging capacity in energy market $Pd_{t,n}$ never exceed its maximum discharging capacity $Pd_{t,n}^{max}$.

\vspace{-3mm}
\subsection{BESS Bidding Model}
% The bidding process consists of two stages: determining day-ahead pre-dispatch bidding prices and adjusting bidding capacities in the real-time market.
\subsubsection{Day-ahead Bidding Model}
%The core of the day-ahead bidding model is to seek the optimal price that the supply from the BESS and other bidders can match the demand.
As required by AEMO, an initial bid with prices and capacities should be submitted 24 hours in advance. Different from the bidding capacities, the submitted bidding prices cannot be further changed.

\begin{subequations}
\begin{small}
\begin{alignat}{2}
\begin{split}
\max &\sum_{\omega=1}^{\Omega} \pi_{\omega} \Bigg\{ \sum_{t=0}^{T} \Big[\sum_{m}(BP_{\omega,t}^{m} BC_{\omega,t,i}^{m}\!-\!M r_{\omega,t}^{+,m}\!-\!M r_{\omega,t}^{-,m}) \\+&p_{\omega,t+1}^e(SoC_{\omega,t+1}-SoC_{\omega,t}) \!-\! \alpha|SoC_{\omega,t+1}\!-\!SoC_{\omega,t}| \Big] \Bigg\}
\end{split}  \label{eq:obj2}\\
\mbox{s.t.}\quad
&\sum_{k=1}^{11}b_{\omega,t,n,k}^m=1, &\label{2b}\\
% &BC_{t,-i}=\sum_{k=2}^{11}b_{t,-i,k}bc_{t,-i,k-1},&\label{3b}\\
&0 \le bc_{t,i,k}^m\le bc_{t,i,k+1}^m\le bc_{t,i}^{max,m},k=1,...,9&\label{2c}\\
&BC_{\omega,t,n}^m=\sum_{k=2}^{11}b_{\omega,t,n,k}^m bc_{t,n,k-1}^m,&\label{2d}\\
% &BP_{t,n}=\sum_{k=1}^{10}b_{t,n,k}bp_{t,n,k},&\label{3e}\\
&0 \le bp_{t,i,k}^m\le bp_{t,i,k+1}^m\le bp^{cap},k=1,...,9&\label{2e}\\
&BP_{\omega,t}^m\ge \sum_{k=2}^{11}b_{\omega,t,n,k}^m bp_{t,n}^m, &\label{2f}\\
&BP_{\omega,t}^m\le \sum_{k=1}^{10}b_{\omega,t,n,k}^m bp_{t,n}^m +b_{\omega,t,n,11}^m bp^{cap}, &\label{2g}\\
&BC_{\omega,t,i}^m+\sum_{-i}BC_{\omega,t,-i}^m+r_{\omega,t}^{+,m}-r_{\omega,t}^{-,m}+s_{\omega,t}^m=d_{\omega,t}^m, &\label{2h}\\
&SoC_{\omega,0}=0.5, &\label{2i}\\
\begin{split}
&SoC_{\omega,t+1}=SoC_{\omega,t}\\&+[(Pc_t+S_t^{\omega,lr}BC_{\omega,t,i}^{lr}+S_{\omega,t}^{lc}BC_{\omega,t,i}^{lc})\eta_c\\&-(Pd_t+S_{\omega,t}^{rr}BC_{\omega,t,i}^{rr}+S_{\omega,t}^{rc}BC_{\omega,t,i}^{rc})/\eta _d]\Delta t/ E_{max}, \end{split}\label{2j}\\
&SoC_{min}\le SoC_{\omega,t} \le SoC_{max}, &\label{2k}\\
&0 \le Pd_{t}\le e_{t}Pd_{t}^{max}, &\label{2l}\\
&0 \le Pc_{t}\le (1-e_{t})Pc_{t}^{max}, &\label{2m}\\
&bc_{t,i,10}^{lr}+bc_{t,i,10}^{lc}+Pc_{t}\le Pc_{t}^{max},& \label{2n}\\
&bc_{t,i,10}^{rr}+bc_{t,i,10}^{rc}+Pd_{t}\le Pd_{t}^{max}, &\label{2o}\\
&r_{\omega,t}^{+,m},r_{\omega,t}^{-,m}\ge 0, &\label{2p}\\
&\alpha=C_{BESS}/[2N(SoC_{max}-SoC_{min}], &\label{2q}\\
&b_{\omega,t,n,k}^m,e_t\in \{0,1\}, &\label{2r}\\
&\forall m\in \{lr,rr,lc,rc\}, &\label{2s}
\end{alignat}
\end{small}
\end{subequations}
where $\pi_{\omega}$ is the probability of scenario $\omega$ in scenario set $\Omega$. The optimization objective \eqref{eq:obj2} is to maximize the profits from the joint energy and FCAS markets with the minimum degradation cost within a 24-hour optimization window $T$. Similar to \eqref{1b}, \eqref{2b} is used for bidding band selection but it seeks the ranges instead of values. Specifically, since the bidding price $BP_{\omega,t}^m$ is variable from 0 to the bidding price cap $bp^{cap}$, it will span 12 prices from 0, ten bidding prices to the price cap and be located in one range from their 11 ranges and that is the reason for 11 binaries. \eqref{2c} and \eqref{2e} ensure that the bidding prices and capacities of the model are monotonically increasing. \eqref{2f}-\eqref{2g} find the bidding bands in all bids in which $BP_{\omega,t}^m$ is located and \eqref{2d} finds the capacities $BC_{\omega,t,n}^m$ in the selected bidding bands. \eqref{2h} ensures that the capacity of the BESS $BC_{\omega,t,i}^m$ plus the capacities of other bidders in market $m$ plus the overall market supply variation $s_t$ matches the demand $d_t$ in market $m$, where $r_{\omega,t}^{+,m}$ and $r_{\omega,t}^{-,m}$ are relaxation variables. \eqref{2i}-\eqref{2k} updates the state of charge (SoC) $SoC_{\omega,t}$ of the BESS within the SoC operation ranges, where $S_t^m\in (0,1)$ is the FR command in market $m$. \eqref{2l}-\eqref{2m} ensure that the charging power $Pc_t$ and the discharging power $Pd_t$ of the BESS are within the maximum charging capacity $Pc_t^{max}$ and the maximum discharging capacity $Pd_t^{max}$ and it cannot be charged and discharged simultaneously. Similar to \eqref{1j}-\eqref{1k}, \eqref{2n}-\eqref{2o} ensure the joint capacities in energy and FCAS markets do not exceed the maximum capacities. \eqref{2q} calculates the coefficient $\alpha$ to estimate the degradation cost of the BESS, where $C_{BESS}$ and $N$ are the battery cell cost and the maximum cycle numbers of the BESS \cite{WANG2023121918}. Note that since the bid should be unique, the variables forming the bid ($Pc_t$,$Pd_t$,$bc_{t,i,k}^m$ and $bp_{t,i,k}^m$) are unique solutions in $\Omega$.

\subsubsection{Real-time Bidding Model}
AEMO allows bidders to adjust capacities via re-bid in response to the latest changes in the real-time market, hence the real-time model aims to adjust the bidding capacities under real-time prediction and fixed bidding prices by the day-ahead model.
\vspace{-3mm}

\begin{subequations}
\begin{small}
\begin{alignat}{2}
\begin{split}
%&\max\quad\sum_{t=0}^{\tau} \Big[\sum_{m}(cp_t^{m} BC_{t,i}^{m}) \\&+p_{t+1}^e(SoC_{t+1}-SoC_{t})- \alpha|SoC_{t+1}-SoC_{t}|\Big]
\max&\quad \sum_{m}\big[cp_t^{m} (BC_{t,i}^{m}+\Delta BC_{t,i}^{m})\big] \\&+p_{t+1}^e(SoC_{t+1}-SoC_{t})- \alpha|SoC_{t+1}-SoC_{t}|
\end{split}  \label{eq:obj3}\\
\mbox{s.t.}\quad
&\sum_{k=1}^{11}b_{t,n,k}^m=1, &\label{3b}\\
&0 \le bc_{t,i,k}^m\le bc_{t,i,k+1}^m\le bc_{t,i}^{max,m},k=1,...,9&\label{3c}\\
&BC_{t,n}^m+\Delta BC_{t,i}^{m}=\sum_{k=2}^{11}b_{t,n,k}^m bc_{t,n,k-1}^m,&\label{3d}\\
&cp_t^{m}\ge \sum_{k=2}^{11}b_{t,n,k}^m bp_{t,n}^m, &\label{3e}\\
&cp_t^{m}\le \sum_{k=1}^{10}b_{t,n,k}^m bp_{t,n}^m +b_{t,n,11}^m bp^{cap}, &\label{3f}\\
&BC_{t,i}^m+\Delta BC_{t,i}^{m}+\sum_{-i}BC_{t,-i}^m+s_{t}^m\ge d_{t}^m, &\label{3g}\\
%&SoC_{0}=0.5, &\label{3h}\\
\begin{split}
&SoC_{t+1}=SoC_t+\big\{[Pc_t+\Delta Pc_t+S_t^{lr}(BC_{t,i}^{lr}+\Delta BC_{t,i}^{lr})\\&+S_t^{lc}(BC_{t,i}^{lc}+\Delta BC_{t,i}^{lc})]\eta_c -[Pd_t+\Delta Pd_t+S_t^{rr}(BC_{t,i}^{rr}\\&+\Delta BC_{t,i}^{rr})+S_t^{rc}(BC_{t,i}^{rc}+\Delta BC_{t,i}^{rc})]/\eta _d\big\}\Delta t/ E_{max}, \end{split}\label{3h}\\
&SoC_{min}\le SoC_t \le SoC_{max}, &\label{3i}\\
&0 \le Pd_{t}+\Delta Pd_{t}\le e_{t}Pd_{t}^{max}, &\label{3j}\\
&0 \le Pc_{t}+\Delta Pc_{t}\le (1-e_{t})Pc_{t}^{max}, &\label{3k}\\
&bc_{t,i,10}^{lr}+bc_{t,i,10}^{lc}+Pc_{t}+\Delta Pc_{t}\le Pc_{t}^{max},& \label{3l}\\
&bc_{t,i,10}^{rr}+bc_{t,i,10}^{rc}+Pd_{t}+\Delta Pd_{t}\le Pd_{t}^{max}, &\label{3m}\\
%&\alpha=C_{BESS}/[2N(SoC_{max}-SoC_{min}], &\label{3n}\\
&b_{t,n,k}^m,e_t\in \{0,1\}, &\label{3n}\\
&\forall m\in \{lr,rr,lc,rc\}, &\label{3o}
\end{alignat}
\end{small}
\end{subequations}
where \eqref{eq:obj3} is to determine the adjusted capacities ($\Delta Pd_{t}$,$\Delta Pc_{t}$ and $BC_{t,i}^{m}$) in the energy and FCAS markets within a 5-minute rolling window. Unlike the rough day-ahead forecasts of the 24-hour stochastic scenarios, the real-time model uses more accurate real-time deterministic forecasts as input parameters. Note that \eqref{3e}-\eqref{3f} aim to seek the located bidding bands of the clearing prices $cp_t^{m}$  instead of the bidding prices $bp_{t,i,k}^{m}$ in the day-ahead model because $bp_{t,i,k}^{m}$ are considered as fixed parameters in the real-time stage. Also, the capacity adjustment in \eqref{3g} should ensure that the supply meets the demand in real time.
\vspace{-3mm}

\subsection{Supply Estimation Model}
Before using the above models, it is necessary to present the model that calculates the overall market supply variation mentioned in the previous models, i.e. $s_t^m$. The clearing price is determined by both the FCAS supply and demand, taking a real-world scenario in the FCAS market as an example, bidders typically put more capacity into arbitrage during peak energy prices, and hence the clearing prices in FCAS markets will be higher to purchase capacity that can meet the demand. However, the data on clearing prices and demands are given by AEMO whereas the data on supply is incomplete. Hence, it is hard to describe the above scenarios caused by the market supply variation. To better describe the relationship between price and supply-demand balance under actual scenarios, we build a model to estimate the overall market supply variation $s_t^m$ under given clearing prices $cp_t^m$ and demands $d_t^m$.

\begin{subequations}
\begin{small}
\begin{alignat}{2}
\begin{split}
&\min\quad\sum_{m}\sum_{t=0}^{T}(s_t^{+,m}+s_t^{-,m})
\end{split}  \label{eq:obj4}\\
\mbox{s.t.}\quad
&\sum_{k=1}^{11}z_{t,n,k}^m=1, &\label{4b}\\
&ec_{t,n}^m=\sum_{k=2}^{11}z_{t,n,k}^m bc_{t,n,k-1}^m,&\label{4c}\\
&cp_t^{m}\ge \sum_{k=2}^{11}z_{t,n,k}^m bp_{t,n}^m, &\label{4d}\\
&cp_t^{m}\le \sum_{k=1}^{10}z_{t,n,k}^m bp_{t,n}^m +z_{t,n,11}^m bp^{cap}, &\label{4e}\\
&\sum_{n=1}^N ec_{t,n}^m+s_t^{m}=d_t^m, &\label{4f}\\
&s_t^m=s_t^{+,m}-s_t^{-,m}, &\label{4g}\\
&s_t^{+,m},s_t^{-,m}\ge 0, &\label{4h}\\
&z_{t,n,k}^m \in \{0,1\}, &\label{4i}\\
&\forall m\in \{lr,rr,lc,rc\}, &\label{4j}
\end{alignat}
\end{small}
\end{subequations}
where the optimization objective \eqref{eq:obj4} is to find the supply variation $s_t^m$ at the supply-demand balancing point. \eqref{4b}-\eqref{4e} can find the available capacity in market $m$ under the given price $cp_t$. Since $cp_t$ and $d_t$ are given, the only unknown variable $s_t^m$ can be calculated by \eqref{4f}-\eqref{4h}. $s_t^m$ can be both positive and negative, which can be divided into a positive component $s_t^{+,m}$ and a negative component $-s_t^{-,m}$.
\vspace{-1mm}

\subsection{Optimization Framework}
\begin{figure}[htbp]
\centerline{\includegraphics[width=0.9\linewidth]{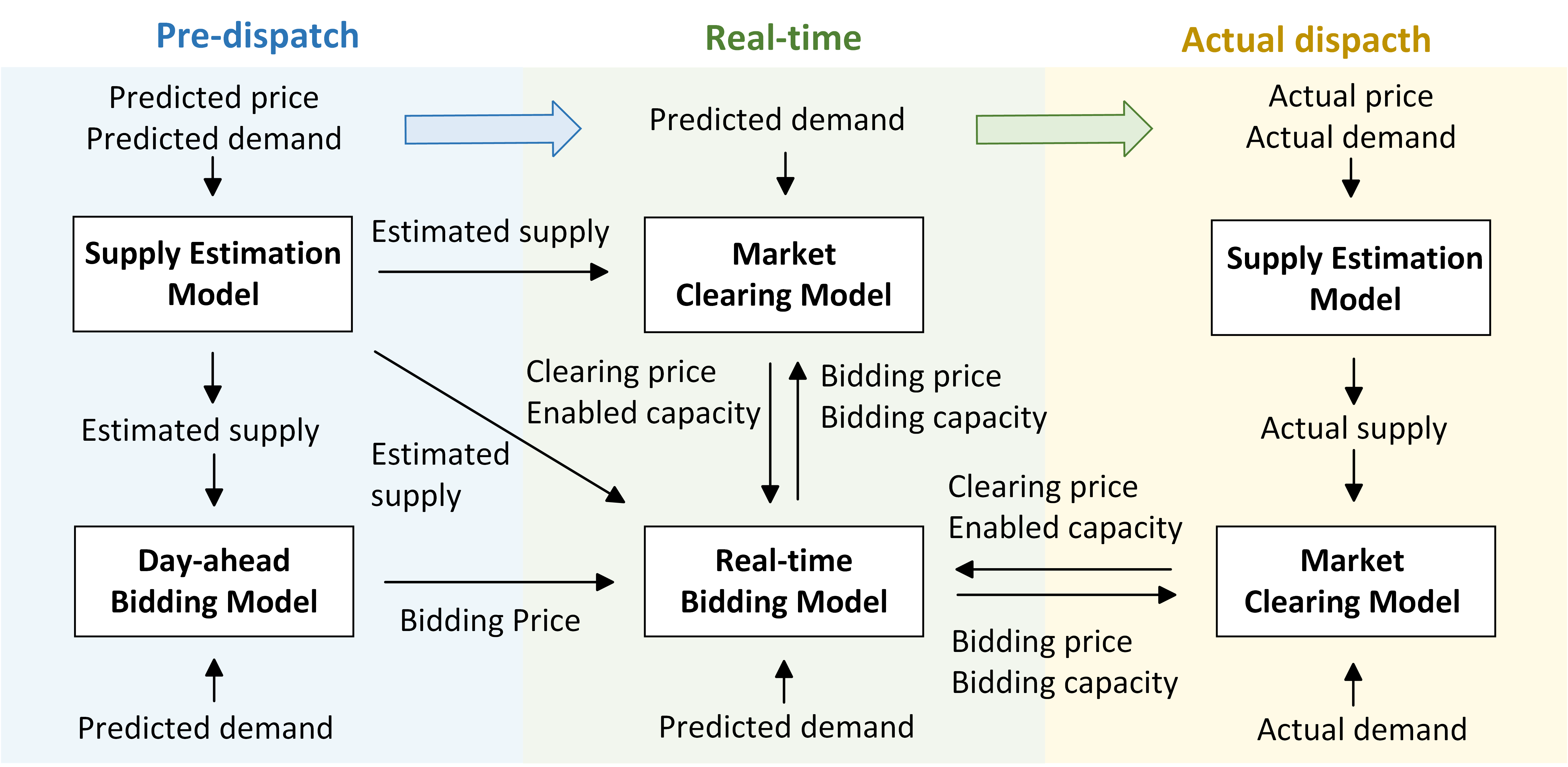}}
\caption{Optimization framework of the proposed models.}
\label{fig:2}
\end{figure}

As shown in Fig.~\ref{fig:2}, we formulate the problem as a bi-level game model, where the upper-level model is the market clearing model as a leader, and the lower-level model is the BESS bidding model as a follower. First, the unknown parameter $s_t^m$ in the problem can be obtained based on the predicted clearing price and demand by solving the supply estimation model \eqref{eq:obj4}. Then, in the pre-dispatch stage, the bidding price and capacity of the BESS are obtained by solving the day-ahead bidding model \eqref{eq:obj2}. In the real-time stage, the bidding price and capacity of the BESS are introduced into model \eqref{eq:obj1} to obtain an initial clearing price. Then, the clearing price is introduced into model \eqref{eq:obj3} to obtain the bidding capacity. In the two-level game, model \eqref{eq:obj1} and model \eqref{eq:obj3} iteratively update their strategies until Nash equilibrium or exceed the iteration limit. Note that all the above models take the predicted market information as inputs. Finally, in the actual dispatch stage, model \eqref{eq:obj4} is solved based on the actual clearing price and demand to obtain the parameter $s_t^m$, and then the bidding strategy from model \eqref{eq:obj3} is introduced into model \eqref{eq:obj1} to obtain the actual clearing price and enabled capacity, as well as the final actual profit of the BESS.

\vspace{-4mm}
\section{Methodology}
%In this section, the MDP formulation of the proposed problem, the LLMs-based decision-making framework and the DRL algorithm are introduced.

\subsection{MDP Formulation}

%RL is a popular machine learning paradigm that can search for optimal strategies during interaction with the environment. %Essentially, RL is the process by which an agent searches a series of decisions to maximize the cumulative reward by interacting with the environment. 
To formulate the proposed problem as a DRL problem, a Markov Decision Process (MDP) including a state space $\mathcal S$, an action space $\mathcal A$, a transition probability function $\mathcal P$, and a reward function $\mathcal R$ is necessary to be defined in the form of a DRL environment for the interaction with the DRL agent.
\begin{itemize}
\item State space: A state space contains a set of states $s_t\in\mathcal S$ that the agent can observe from the environment. At time $t$, the SoC of BESS $SoC_t$, the predicted energy price $p_{t}^e$, FCAS prices $p_{t}^{m}$ and demands $d_{t}^{m}$ in market $m$ are defined as states.
\item Action space: An action space contains a set of actions $a_t\in\mathcal A$ which is the bidding strategy of the agent. At time $t$, the bidding capacities $bc_{t,k}^{lr},bc_{t,k}^{rr},bc_{t,k}^{lc},bc_{t,k}^{rc}, k=1,...,9$ in the FCAS markets as well as the output $Pc_t$ and $Pd_t$ in the energy market are defined as actions.
\item Transition probability function: $\mathcal P(s_t,a_t,s_{t+1})=\Pr(s_{t+1}\mid s_t,a_t)$. The state $s_{t+1}$ at $t+1$ is uncertain under given observation $s_t$ and bidding strategy $a_t$ at $t$.
\item Reward function: $\mathcal S\times\mathcal A\rightarrow  \mathbb R$. When the agent takes an action $a_t$ and transforms $s_t$ to $s_{t+1}$, it receives a reward $r_t (s_t,a_t,s_{t+1})\in \mathcal R$. We define the profits of the BESS including the revenues in the joint energy and FCAS markets and the degradation costs as the rewards. Similar to the objective function of model \eqref{eq:obj3}, the basic reward function can be written as:
\begin{equation}
\begin{split}
r_{t+1}=\sum_{m}(cp_t^{m} ec_{t,i}^{m}) +p_{t+1}^e(SoC_{t+1}-SoC_{t})\\- \alpha|SoC_{t+1}-SoC_{t}| \label{eq:14}
\end{split}
\end{equation}
where $cp_t^{m}$ and $ec_{t,i}^{m}$ are the solutions of the market clearing model \eqref{eq:obj1}, which means that there are interactions between the agent and the market. 
\end{itemize}
Unlike the instantaneous rewards in FCAS, arbitrage rewards tend to be sparse and delayed. To encourage the BESS to charge before discharging and receiving rewards, the potential-based reward shaping (PBRS) \cite{Ng1999PolicyIU} is employed to assist the agent in learning strategies for joint markets. The reward function \eqref{eq:14} is modified by adding an extra reward based on prior knowledge:
\begin{equation}
r_{t+1}\leftarrow r_{t+1}+\gamma\Phi(s_{t+1})-\Phi(s_{t})\label{eq:14-1}
\end{equation}
\begin{equation}
\Phi(s_{t})=\big[\textstyle\sum_{t=0}^{t}(Pc_t\eta_c\Delta t) - \underline{E}\big]^- \label{eq:14-2}
\end{equation}
where potential function $\Phi(s_{t})$ measures the gap between charged energy and energy threshold $\underline{E}$. $[\bullet]^-$ is the projector of $\min(\bullet,0)$ to avoid overcharging. In general, the dense artificial reward signals drive the agent to reach the sub-goal of charging at least a certain amount of energy in each episode.

In order to ensure that the bidding strategy of the agent complies with market standards and rules, any action out of specification is limited, for $k=1,...,9$ we have:
\begin{equation}
bc_{t,k+1}^{lr}\leftarrow\operatorname{clip}(bc_{t,k+1}^{lr},bc_{t,k}^{lr},Pc_t^{max}\!-\!Pc_t)
\label{eq:15}
\end{equation}
\begin{equation}
bc_{t,k+1}^{lc}\leftarrow\operatorname{clip}(bc_{t,k+1}^{lc},bc_{t,k}^{lc},Pc_t^{max}\!-\!Pc_t\!-bc_{t,10}^{lr})
\label{eq:16}
\end{equation}
\begin{equation}
bc_{t,k+1}^{rr}\leftarrow\operatorname{clip}(bc_{t,k+1}^{rr},bc_{t,k}^{rr},Pd_t^{max}\!-\!Pd_t)
\label{eq:17}
\end{equation}
\begin{equation}
bc_{t,k+1}^{rc}\leftarrow\operatorname{clip}(bc_{t,k+1}^{rc},bc_{t,k}^{rc},Pd_t^{max}\!-\!Pd_t\!-bc_{t,10}^{rr})
\label{eq:18}
\end{equation}
where the maximum function is used to ensure the bidding capacities are monotonically increasing in each FCAS market. The minimum function is used to ensure the joint capacities in energy and FCAS markets are within the limitations of the maximum charging or discharging capacity. In addition, when $SoC_t$ is larger than $SoC_{max}$, $Pc_t,bc_{t,k}^{lr},bc_{t,k}^{lc},k=1,...,10$ are limited to 0, and when $SoC_t$ is smaller than $SoC_{min}$, $Pd_t,bc_{t,k}^{rr},bc_{t,k}^{rc},k=1,...,10$ are limited to 0, which ensures $SoC_t$ is always within the limitations of the SoC. Once the MDP is defined, the DRL agent will interact with the environment to obtain experience samples and learn strategies to maximize the cumulative rewards of the scenarios. After offline training, the trained model will be applied to the online bidding system as the decision engine.

\subsection{LLMs-assisted AI-agent Interactive Decision-making Framework}
The epistemic uncertainty leads to a declined bidding performance of the DRL agent in unfamiliar scenarios outside the training set. Thus, we propose an LLMs-assisted AI-agent interactive decision-making framework to decompose the complicated bidding task into more manageable sub-tasks and deploy multiple LLMs-based agents to address the components in an orderly modular pattern. 
%The outcomes of subtasks will engage in the re-shape of MDP that drives the DRL agent to adapt to the new environment and deliver feedback in a loop as interpreters to reinforce the LLMs-assisted agents. Through the interactive collaboration between the DRL agent and the LLMs-assisted agents, the strategy timeliness, reliability, and interpretability in new scenarios during the online operation will be enhanced.
Compared to the DRL model trained on historical data, LLMs can analyze the latest market conditions in real time, enhancing the strategy timeliness. To reduce profit loss in unfamiliar scenarios, LLMs can adjust inferior strategies by conditional hybrid decisions and incrementally integrate new knowledge into DRL through MDP reshaping, thereby improving the strategy reliability. Additionally, unlike the "black box" nature of DRL networks, LLMs can act as interpreters to provide explanations to users, thereby increasing the interpretability. An external evaluation and self-reflection mechanism are designed in the “DRL-LLMs” loop to mitigate the hallucination problem in LLMs.

\subsubsection{Definition of LLMs-assisted Agents}
\begin{itemize}
\item Data Analyst (Agent-1): Agent-1 aims to detect whether the state space $\mathcal S$ during online operation is common in the historical training set (\textbf{True}) or not (\textbf{False}). Then it analyzes the potential profitability in each FCAS market from historical statistics, predicted price trends, and prediction quality for subsequent adjustment guidance.

% At each interval, the agent executes the numerical statistics of the historical market data and analyzes the prediction information and the bidding strategies. Then, the agent will determine the profitability indicators reflecting the confidence degree in profitability (-1: low profitability; 0: medium profitability; 1: high profitability). These indicators will be added in the state space $\mathcal S$ so that the DRL agent can observe the analyzed results by LLMs as extra observed information even if facing new data outside the training dataset. 

\item Strategy Optimizer (Agent-2): The target of Agent-2 is to adopt a hybrid decision $a_t'$ to make higher profits $r_t'$:
\begin{equation}
a_t'=a_t+\Delta a_t^{LLMs}\textbf{1}(\textbf{False}\land \textstyle\sum_{\tau=t-h}^{t-1} r_\tau'\ge \sum_{\tau=t-h}^{t-1} r_\tau)  \label{eq:a2}
\end{equation}
To mitigate unreliable decisions by LLMs' hallucination, we introduce another condition as an external evaluation mechanism to constrain $a_t'$ rigidly. $a_t'$ is enabled only when its cumulative return within a half-hour rolling horizon $h$ is higher than that of $a_t$, thereby ensuring a safe profit margin. Dynamically varying conditions over time also avoid a consistent bias towards one alternative.

% The profitability indicators of Agent-1 will contribute to the decision-making of Agent-2 and the bidding strategy from Agent-2 will also be considered for the indicator determination of Agent-1.
\item Performance Evaluator (Agent-3): With the ability of LLMs' self-reflection and multi-agent interaction, Agent-3 evaluates the bidding performance of $a_t$ and $a_t'$ and then provides feedback to reinforce the agents. It can also act as an interpreter for agents' decision explanations.

% Based on the expected and actual profits in the FCAS markets, Agent-3 will evaluate and explain if the profitability indicators are reasonable and generate its evaluation report as feedback to Agent-1. Based on the bidding result of the selected model by Agent-2, Agent-3 will evaluate and explain its performance and generate its evaluation report as feedback to Agent-2. In addition, Agent-3 can help users understand the agents' behavior and motivation, and enhance the interpretability of the bidding process.
\end{itemize}

\subsubsection{Interactive Agent Communication and Collaboration}
%To utilize LLMs to play different roles in the framework, 
We first define the user prompt $\mathbb P$ and formalize it as $\mathbb P=(\mathbb O,\mathbb A,\mathbb R,\mathbb C)$ which includes observation $\mathbb O$, action $\mathbb A$, reward $\mathbb R$ and context $\mathbb C$ to formulate the problem for LLMs. 
\begin{itemize}
\item Observation $\mathbb O$ contains the state space $\mathcal S$, the action space $\mathcal A$, and the market historical statistics and prediction. 
\item Action $\mathbb A$ needs to be defined individually for each agent. For Agent-1, the actions are the detected and analyzed results on $\mathbb O$ in text form. For Agent-2, the actions are the adjusted bidding capacities in numerical form. For Agent-3, the actions are the evaluation results in text form.
\item Reward $\mathbb R$ contains the numerical performance of the decisions, which will be converted to text form as feedback.
\item Context $\mathbb C$ includes the problem description, the task definitions, the task targets, the constraints on action $\mathbb A$, the expected answer format, and any necessary information.
\end{itemize}

For a given prompt $\mathbb P$, LLMs can observe $\mathbb O$ and determine the agents' actions $\mathbb A$ in the requested format. The multi-agent workflow is organized as follows. First, Agent-1 sends its detected and analyzed results to Agent-2. Then Agent-2 performs hybrid decisions based on the current conditions and the given information. The original $a_t$ and the adjusted $a_t'$ are executed synchronously in the environment to obtain the rewards $r_t$ and $r_t'$. Finally, Agent-3 provides feedback including quantitative assessments and qualitative analysis based on each agent's performance for strategy correction at subsequent iterations.

During online operation, LLMs can realize interactive collaboration between multiple LLMs-assisted AI agents and the DRL agents by affecting the MDP in the environment. First, the tuple ($s_t$,$a_t$,$r_t$) under the original strategy will be replaced by the new tuple ($s_t'$,$a_t'$,$r_t'$) under the LLMs-assisted new strategy and stored in the experience replay. Then the DRL network parameters are fine-tuned based on gradients through incremental learning from superior trajectories. Unlike the traditional method of penalizing inferior trajectories, adjusted trajectories can mitigate the profit loss in unfamiliar scenarios and drive DRL to better adapt to new environments.

\subsection{CVaR-DRL Algorithm}
In addition to epistemic uncertainty, aleatoric uncertainty also impacts the profitability of the BESS. Due to the prediction errors, the observation of DRL agent exists certain noises in our problem instead of fixed in some traditional problems. To reduce the risk of uncertain MDP and improve the algorithm performance, the CVaR-DRL algorithm is proposed, where the proximal policy optimization (PPO) architecture is employed as the basic decision-making engine, CVaR \cite{ying2022towards} is introduced as risk metrics and the maximum entropy principle \cite{eysenbach2021maximum} is applied for model robustness enhancement. 

A standard DRL aims to find a policy $\pi_\theta$ parameterized by $\theta$ that maximizes the expected cumulative reward:
\begin{equation}
\operatorname{argmax}\mathbb E_{\pi_\theta}\left [\sum_{t=0}^{\infty } \gamma^t r_t(s_t,a_t)  \right ]   \label{eq:19}
\end{equation}
where $\gamma\in(0,1)$ is a discount factor. Deterministic policy often struggles to cope with uncertainty in MDPs and tends to fall into local optimum, hence, we first combine the concept of entropy with DRL to introduce randomness into the policy so that DRL agents can fully explore the environment. Mathematically, the objective function can be modified as follows:
\begin{equation}
J(\pi_\theta) = \mathbb E_{\pi_\theta} \bigg\{ R(\pi_\theta)\triangleq\!\sum_{t=0}^{\infty } \gamma^t \Big[ r_t(s_t,a_t)+\delta\mathcal H_{\pi_\theta}(\pi_\theta(a_t\mid s_t))\Big] \!\bigg\}   \label{eq:191}
\end{equation}
where the term $\mathcal H_{\pi_\theta}$ maximizes the policy entropy under coefficient $\delta$ to encourage the agent to adopt the action distributions with high entropy and improve the generalization capabilities of policy under the uncertainty. 

In addition, we introduce CVaR to quantify the expected loss under uncertainty in the worst $1-\alpha$ scenarios:
\begin{equation}
\operatorname{CVaR}_{\alpha}\big(R(\pi_\theta)\big) = \min_{\mu} \bigg\{ \mu + \frac{\mathbb E_{\pi_\theta}  \big[R(\pi_\theta)-\mu \big]^+}{1-\alpha} \!\bigg\}  \label{eq:20}
\end{equation}
where $\alpha\in(0,1)$ is the confidence level. $[\bullet]^+$ denotes the projector of $\max(\bullet,0)$. $\mu$ denotes the threshold in the loss distribution. $R(\pi_\theta)$ is the trajectory return under policy $\pi_\theta$. Based on the optimization method in \cite{chow2014algorithms}, \eqref{eq:20} can be constrained by CVaR to avoid profit loss in the worst scenarios. The rewritten objective function is as follows:

\begin{equation}
\min_{\pi_\theta} -J(\pi_\theta) \quad\operatorname{s.t.} \operatorname{CVaR}_{\alpha}\big(-R(\pi_\theta)\big)\le \beta  \label{eq:21}
\end{equation}
where $\beta$ is the reward tolerance. Then the Lagrangian relaxation is employed to unconstrain the problem:

\begin{equation}
\max_{\lambda}\min_{\theta,\mu} L(\lambda,\theta,\mu)\!\triangleq\!-J(\pi_\theta) \!+\!\lambda \bigg\{\beta\!-\! \mu \!+\! \frac{\mathbb E_{\pi_\theta}  \big[\mu-R(\pi_\theta) \big]^+}{1-\alpha}\bigg\}   \label{eq:200}
\end{equation}
where $\lambda$ is the Lagrange multiplier. Based on the PPO architecture, the objective function \eqref{eq:200} can be optimized by the following gradients \eqref{eq:201}-\eqref{eq:203} in terms of $\lambda$, $\theta$ and $\mu$:

\begin{equation}
\nabla_\lambda L(\lambda,\theta,\mu)= \beta- \mu + \frac{\mathbb E_{\pi_\theta}  \big[\mu-R(\omega_{\pi_\theta} \big]^+}{1-\alpha}   \label{eq:201}
\end{equation}
\begin{equation}
\begin{aligned}
&\nabla_\theta L(\lambda,\theta,\mu)\!=\\& \!\nabla_\theta\min \Big\{\frac{\pi_{\theta}(a \mid s)}{\pi_{\theta_{k}}(a \mid s)} \hat{A}^{\pi_{\theta_{k}}},  
\operatorname{clip}\big[\frac{\pi_{\theta}(a \mid s)}{\pi_{\theta_{k}}(a \mid s)}, 1-\epsilon, 1+\epsilon\big] \hat{A}^{\pi_{\theta_{k}}}\Big\}\\&
-\mathbb E_{\pi_\theta} \big\{\nabla_\theta \operatorname{log} P_\theta[R(\omega_{\pi_\theta})]\big\} \Big\{ R(\omega_{\pi_\theta} \!-\! \frac{\lambda \mathbb E_{\pi_\theta}\big[\mu-R(\omega_{\pi_\theta} \big]^+}{1-\alpha} \Big\}  \label{eq:202}
\end{aligned}
\end{equation}
\begin{equation}
\nabla_\mu L(\lambda,\theta,\mu)= \frac{\lambda \mathbb E_{\pi_\theta}\textbf{1}  \big[ \mu\ge R(\omega_{\pi_\theta}) \big]}{1-\alpha} - \lambda  \label{eq:203}
\end{equation}
where $\omega_{\pi_\theta}$ is the scenario under policy $\pi_\theta$. $\textbf{1}[\bullet]$ is the indicator function. $\hat{A}^{\pi_{\theta_{k}}}$ is the estimated advantage function, $\frac{\pi_{\theta}(a \mid s)}{\pi_{\theta_{k}}(a \mid s)}$ is the importance sampling ratio, $\epsilon$ is the clip parameter and the clip function $\operatorname{clip}(\cdot)$ is used to remove the overestimation and underestimation by upper limit $1+\epsilon$ and lower limit $1-\epsilon$. Typically, the advantage function is estimated by the Generalized Advantage Estimator (GAE):
\begin{equation}
\hat{A}^{\pi_{\theta_{k}}} = \sum_{t=0}^{\infty }(\gamma\nu)^{t'}\big[r_{t+t'}+\gamma V(s_{t+t'+1})-V(s_{t+t'})\big]
\label{eq:22}
\end{equation}
where $\nu$ is the GAE parameter and V(s) is the value function. 

In summary, Algorithm 1 presents the implementation and procedure of the LLMs-assisted AI-agent CVaR-DRL.

\begin{algorithm}[htbp]
\begin{small}
\caption{LLMs-assisted AI-agent CVaR-DRL.}
\LinesNumbered
Parameter initialization: policy parameter $\theta$, clip threshold $\epsilon$, GAE parameter $\nu$, entropy coefficient $\delta$, confidence level $\alpha$ and reward tolerance $\beta$\;
% \textbf{Offline Training:}\\
% \For{Episode $n=1,2,3\cdots N$}{
% Randomly sample a scenario $\omega$ and reset the SoC of the BESS\;
% \For{$t=1,2,3\cdots T$}{
% DRL agent generates bidding strategy $a_t$ based on observed state $s_t$\;
% Compute reward $r_t$ and update next state $s_{t+1}$\;
% }
% Collect trajectories under DRL agent's strategy $\tau=(s_1,a_1,r_1,s_2,a_2,r_2,...,s_T,a_T,r_T)$\;
% }
% \textbf{Online Operation:}\\
\For{Episode $e=1,2,3\cdots E$}{
\For{$t=1,2,3\cdots T$}{
DRL agent observes state $s_t$ and generates the bidding strategy $a_t$\;
Agent-1 sends detected results of $\mathcal S$ and analyzed market profitability to Agent-2\;
Agent-2 performs hybrid decisions $a'_t$ based on the current conditions and the given information\;
The market clearing model computes the rewards $r_t$ and $r'_t$ based on $a_t$ and $a'_t$ respectively\;
Agent-3 evaluates $r_t$ and $r'_t$ and generates feedback to Agent-1 and Agent-2\;
Update the state to $s'_{t+1}$ based on $a'_t$\;
}
Collect trajectories under the LLMs-assisted strategy $\tau=(s_1,a'_1,r'_1,s'_2,a'_2,r'_2,...,s'_T,a'_T,r'_T)$\;
}
% \textbf{Policy Update:}\\
\For{Iteration $i=1,2,3\cdots I$}{
Sample $\tau$ from the replay buffer\;
Estimate the advantage $\hat{A}^{\pi_{\theta_k}}$ based on \eqref{eq:22}\;
Update the Lagrange multiplier $\lambda$ based on \eqref{eq:201}\;
Update the policy parameter $\theta$ based on \eqref{eq:202}\;
Update the threshold $\mu$ based on \eqref{eq:203}\;
}
\end{small}
\end{algorithm}

\section{Case Study}
The datasets in this paper are provided by AEMO, including historical clearing prices, demands, and bids in FCAS markets and energy prices in South Australia. ARIMA is employed to predict market prices and demands. The FR commands are simulated based on the regulation D signal from PJM. The scenarios are randomly sampled from the dataset in 24-hour periods. The time interval of the MDP is 5 minutes which matches the trading interval of NEM. The capacity of the BESS is 100MWh with an SoC range from 0.2 to 0.8. The charging/discharging power is 50MW with an efficiency of 0.95\%. The battery cell cost is \$0.5/Wh with a maximum cycle number of 10,000. $\underline{E}$ is set at 40MWh. The value of $\alpha$ is 0.9, determined by identifying the inflection point on the CVaR curve with respect to $\alpha$. The value of $\beta$ is 12,500, calculated as the average of CVaR and VaR at selected $\alpha$. The hyperparameter setting for the DRL algorithm is shown in Table \ref{tab2}. LLMs employ the latest GPT-4 Turbo model.

% \begin{table}[htbp]
% \vspace{-0.2cm}
% \caption{Parameters of the BESS}
% \begin{center}
% \resizebox{1\columnwidth}{!}{
% \begin{tabular}{cccc}
%    \toprule
%    Parameter & Value & Parameter & Value \\
%    \midrule
%    Capacity & 100MWh & Battery cell cost & \$0.5/Wh \\
%    Maximum power & 50MW  & Maximum cycle number & 10,000  \\
%    Maximum SoC & 80\% & Charging efficiency & 0.95\%\\
%    Minimum SoC & 20\% & Discharging efficiency  & 0.95\% \\
%    \bottomrule
% \end{tabular}}
% \label{tab1}
% \end{center}
% \end{table}

\begin{table}[htbp]
\vspace{-0.2cm}
\caption{Hyperparameter setting}
\begin{center}
\resizebox{0.9\columnwidth}{!}{
\begin{tabular}{cccc}
   \toprule
   Hyperparameter & Value & Hyperparameter & Value \\
   \midrule
   Batch size & 2880 & Mini-batch size & 72\\
   Actor learning rate & 0.001  & Critic learning rate & 0.001  \\
   Discount factor & 0.99 & Entropy coefficient & 0.01 \\
   Clip threshold & 0.2 & GAE parameter & 0.95 \\
   Hidden layer & 2 &   Node number & [256,256]\\
   Activation function & Tanh & Optimizer & Adam \\
   \bottomrule
\end{tabular}}
\label{tab2}
\end{center}
\vspace{-0.6cm}
\end{table}

\subsection{Discussion on Market Clearing Model}
\begin{figure}[htbp]
\centerline{\includegraphics[width=0.75\linewidth]{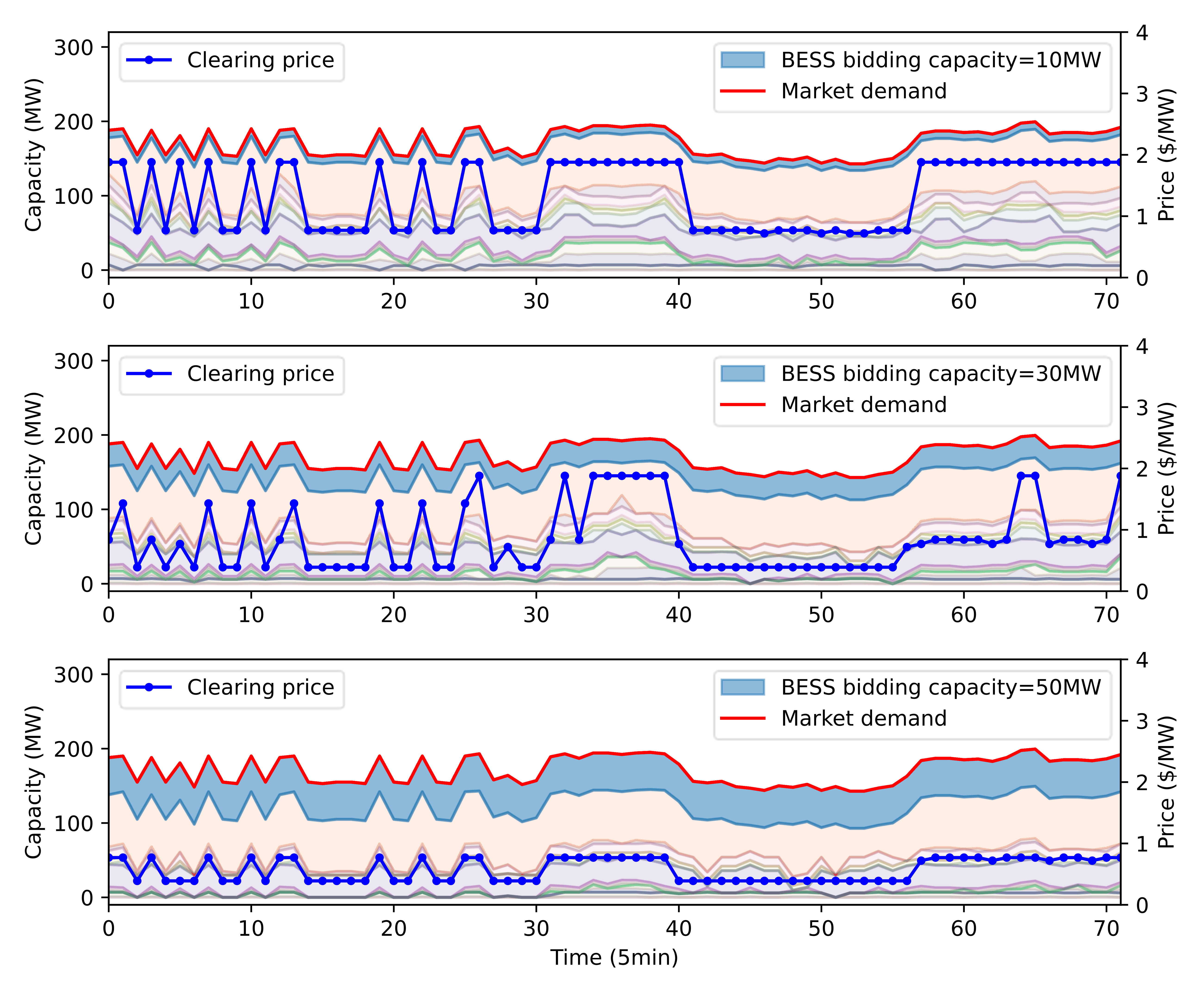}}
\caption{Clearing results of the market clearing model under different BESS bidding capacities.}
\label{fig:r1}
\end{figure}

The role of the market clearing model is to acquire sufficient capacities from various bidders to satisfy the market demand and establish clearing prices. Fig.~\ref{fig:r1} illustrates the clearing results for all participants in the market clearing model when the BESS bids 10MW, 30MW, and 50MW respectively. When the BESS bids 10 MW, the market clearing model has to purchase additional capacities from other bidders to ensure the supply-demand balance, resulting in higher clearing prices. As the bidding capacity of the BESS increases to 30MW, fewer capacities are needed from other bidders, leading to a decrease in clearing prices. As the BESS bids 50MW, the clearing prices are significantly lower. Thus, as a price maker, the BESS's bidding capacities influence the market clearing prices.

\vspace{-0.2cm}

\subsection{BESS Cost-benefit Analysis}
\begin{figure}[htbp]
\centerline{\includegraphics[width=0.95\linewidth]{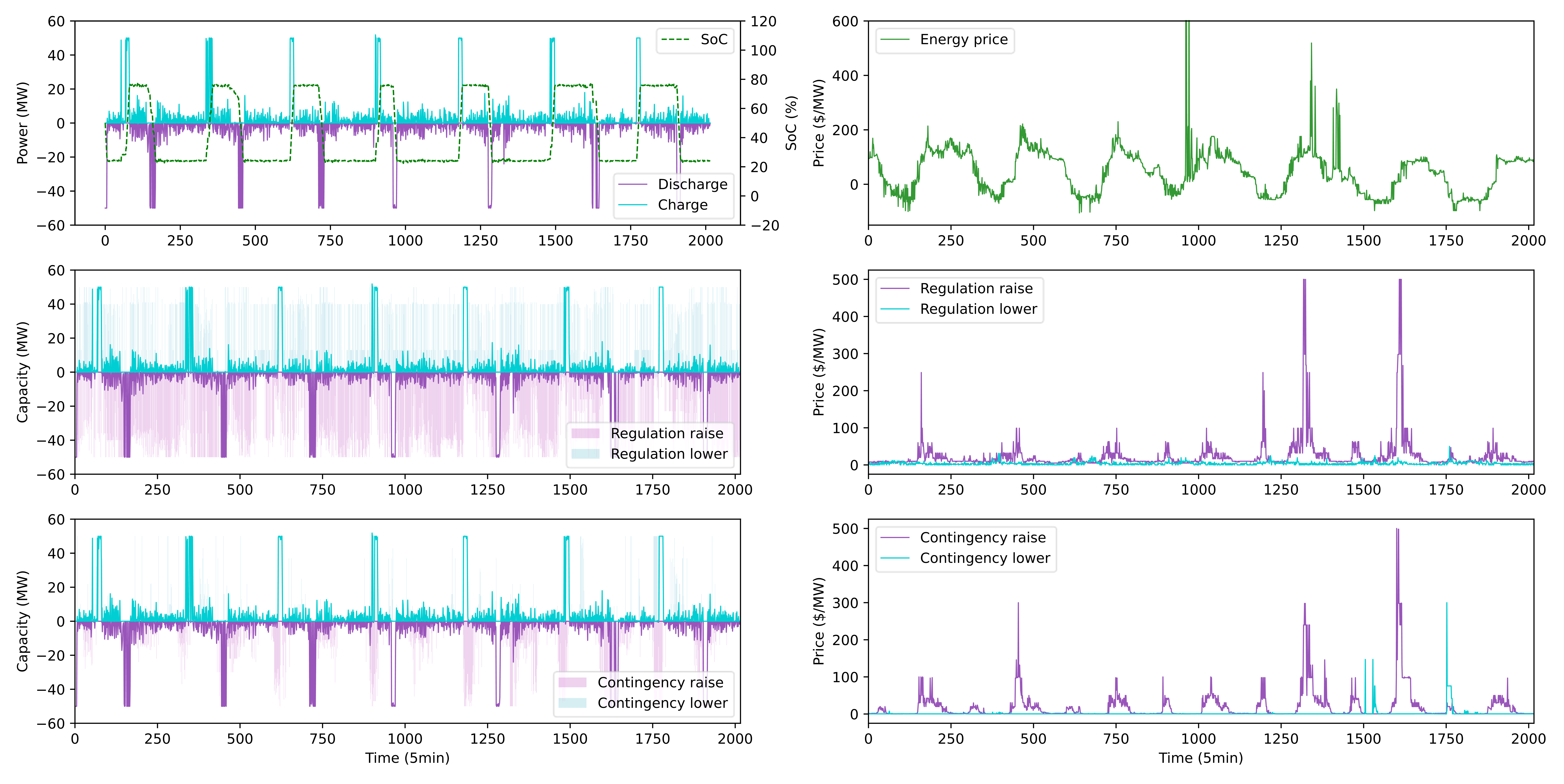}}
\caption{BESS bidding results of the mathematical model.}
\label{fig:r2}
\end{figure}

\begin{figure}[htbp]
\centerline{\includegraphics[width=0.95\linewidth]{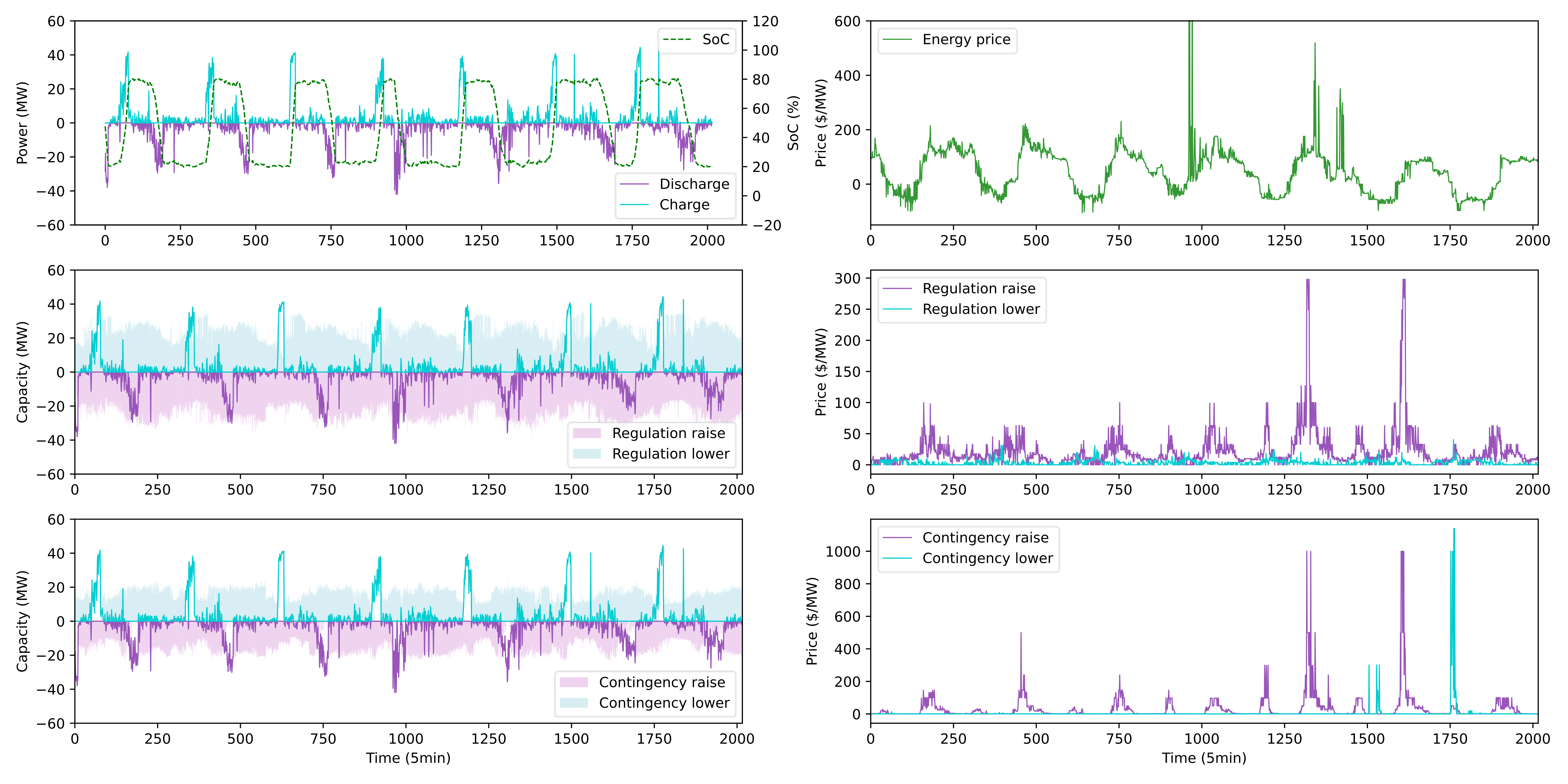}}
\caption{BESS bidding results of the DRL model.}
\label{fig:r3}
\end{figure}

To assess the profitability of the mathematical and DRL models, we evaluate their benefit performance in the joint energy and FCAS markets. Fig.~\ref{fig:r2} displays the bidding results of the mathematical model, wherein the BESS bids higher capacities during price spikes in the contingency markets, while mainly allocating capacities to the regulation markets under normal conditions. Fig.~\ref{fig:r3} shows the bidding results of the DRL model, which tends to distribute its capacities more uniformly across both the contingency and regulation markets.

\begin{figure}[htbp]
\centerline{\includegraphics[width=0.93\linewidth]{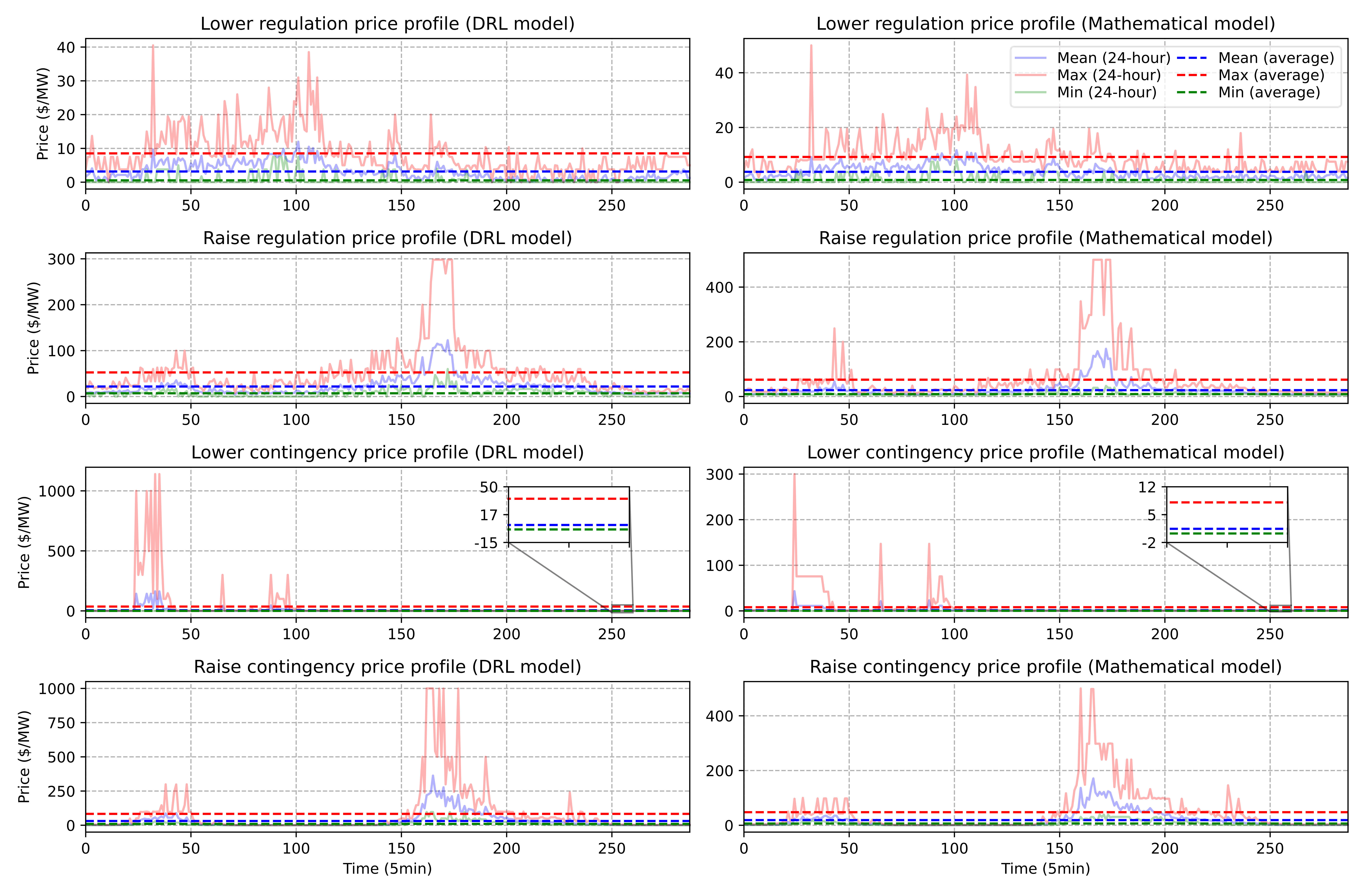}}
\caption{Statistics on clearing prices under the strategies of DRL and mathematical models.}
\label{fig:r4}
\end{figure}

Fig.~\ref{fig:r4} provides statistical data on the clearing prices under both strategies, which aggregates the maximum, mean, and minimum values of each FCAS market into a 24-hour profile, along with their averaged values. Generally, the market trends show that raised markets typically have higher prices than lower markets, and regulation markets are priced above contingency markets. Nevertheless, the contingency markets exhibit greater price volatility, offering rare but large profit opportunities. In both bidding strategies, the models generally allocate more capacities to the regulation market than to the contingency markets to secure stable revenues. However, notable differences exist in handling contingency capacities: the mathematical model tends to shift regulation capacities to the contingency markets after predicting the price spikes, whereas the DRL model consistently bids a portion of capacities into the contingency market to pursue potential profit opportunities. Additionally, the clearing results are different between the models, where the peak prices achieved by the DRL model in both contingency raise and lower exceed \$1,000/MW, significantly surpassing those of the mathematical model.

% \vspace{-0.2cm}

% \vspace{-0.2cm}

% \vspace{-0.2cm}

Table \ref{tab3} provides a summary of the profits of the BESS under the strategies of the DRL model and the mathematical model. In the energy market, the arbitrage profits of the BESS utilizing the DRL model show an increase of 7.69\% over those achieved by the mathematical model. In the regulation markets, profits from the DRL model both lower and raise regulation are increased by 62.23\% and 21.90\%, respectively. Additionally, the DRL model enhances profits in the contingency markets by 8.77\% and 7.67\% and achieves a 5.63\% reduction in degradation costs. Overall, the BESS under the DRL model generates \$37,163 more in profit than the mathematical model, which demonstrates the superior performance of the DRL model in maximizing bidding benefits.

\begin{table}[htbp]
% \vspace{-0.2cm}
\caption{Profit comparison}
\begin{center}
\resizebox{0.9\columnwidth}{!}{
\begin{tabular}{ccc}
   \toprule
   Revenue & DRL model & Mathematical model  \\
   \midrule
   Energy arbitrage & \$81,240 & \$75,436 \\
   Regulation lower & \$13,937 & \$8,591  \\
   Regulation raise & \$98,160  & \$80,525  \\
   Contingency lower & \$6,387 & \$5,872 \\
   Contingency raise & \$73,199 & \$67,985 \\
   Degradation cost & -\$47,024 &  -\$49,673 \\
   \midrule
   Overall profit & \$225.899 &  \$188,736  \\
   \bottomrule
\end{tabular}}
\label{tab3}
\end{center}
\vspace{-0.8cm}
\end{table}

% PLC PRC PLR PRR

% 1.431429401	18.82206317	3.79015377	23.27147645

% 5.417638889	30.81334821	3.163943452	21.85816468

\subsection{Evaluation of Bidding Risk}
\begin{figure}[htbp]
\centerline{\includegraphics[width=0.85\linewidth]{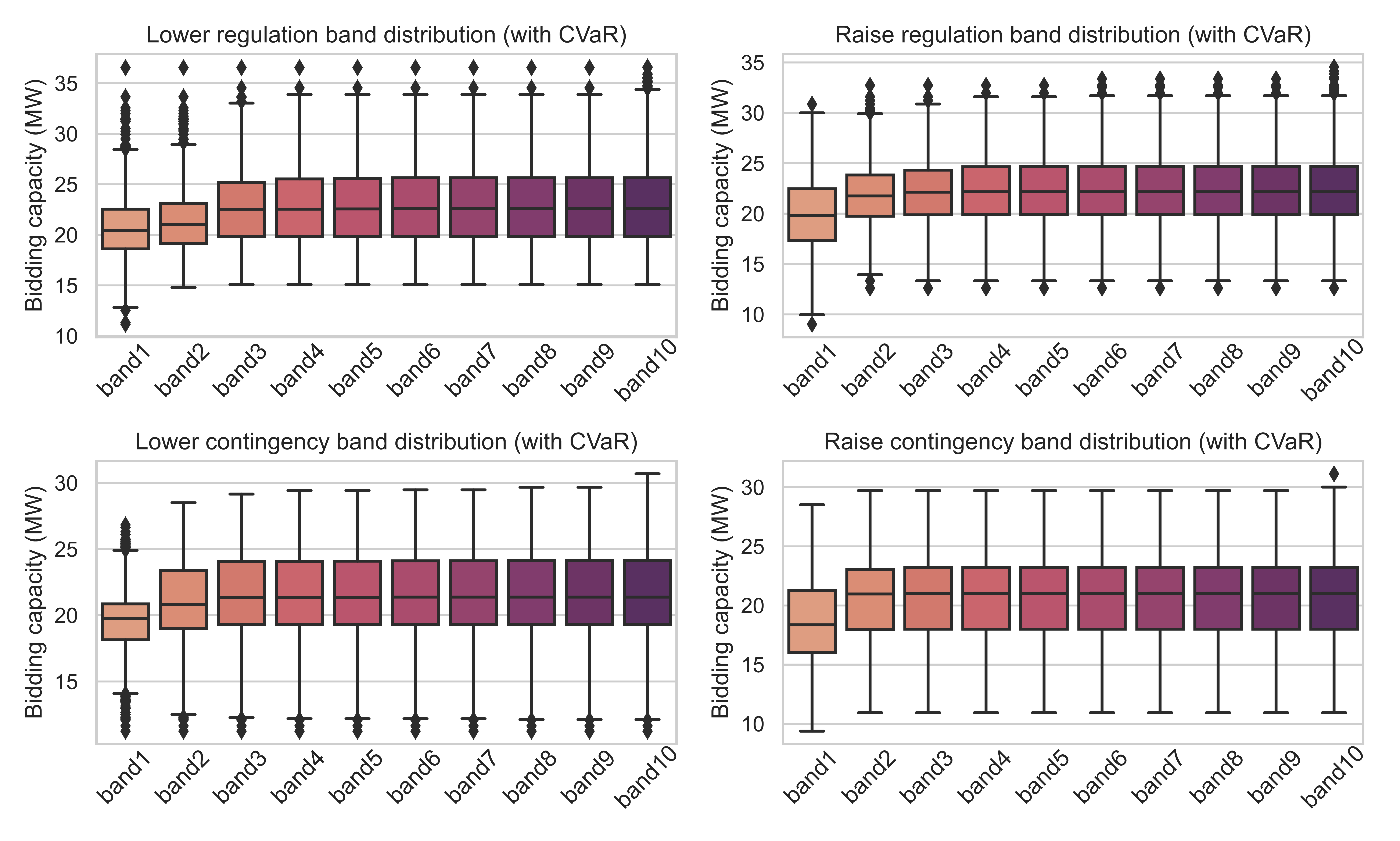}}
\caption{BESS bidding distribution of the CVaR-DRL model.}
\label{fig:r5}
\end{figure}

%From the previous analysis, the DRL approach presents a certain robustness to risk compared with the mathematical approach, which explains why in the presence of forecast uncertainty the DRL agent allocates capacities more evenly between different markets to capture uncertain profit opportunities. 
To assess the role of risk management in bidding strategies, we analyze the detailed distribution of bidding bands of each market for both the CVaR-DRL and the original DRL algorithms. Fig.~\ref{fig:r5} presents the box plot representation of the BESS bidding distribution for the CVaR-DRL model. The bidding bands are monotonically increasing and range from band 1 to band 10. Band1 represents the maximum capacity that can be purchased at the lowest clearing price, whereas band10 corresponds to the maximum capacity at the highest price. A higher concentration of capacity in the lower bands suggests that the BESS prioritizes having its bids accepted and disregards the risk of lower clearing prices and profits. Conversely, a concentration in the higher bands indicates a pursuit of higher clearing prices, with less consideration of the risk of bid failure. In terms of median, bidding capacities tend to be low in band1 across all markets, gradually increasing up to approximately band3, which illustrates the BESS attempts to elevate clearing prices while maintaining stable revenue. Beyond band 3, the lower extremes of capacity bids stabilize above 10MW, with the upper extremes continuing to rise slightly, indicating that in some scenarios, the BESS might incrementally increase bids to between 30MW and 35MW to further elevate the clearing prices. The outliers highlight the capacity distribution in rare cases.

\begin{figure}[htbp]
\centerline{\includegraphics[width=0.85\linewidth]{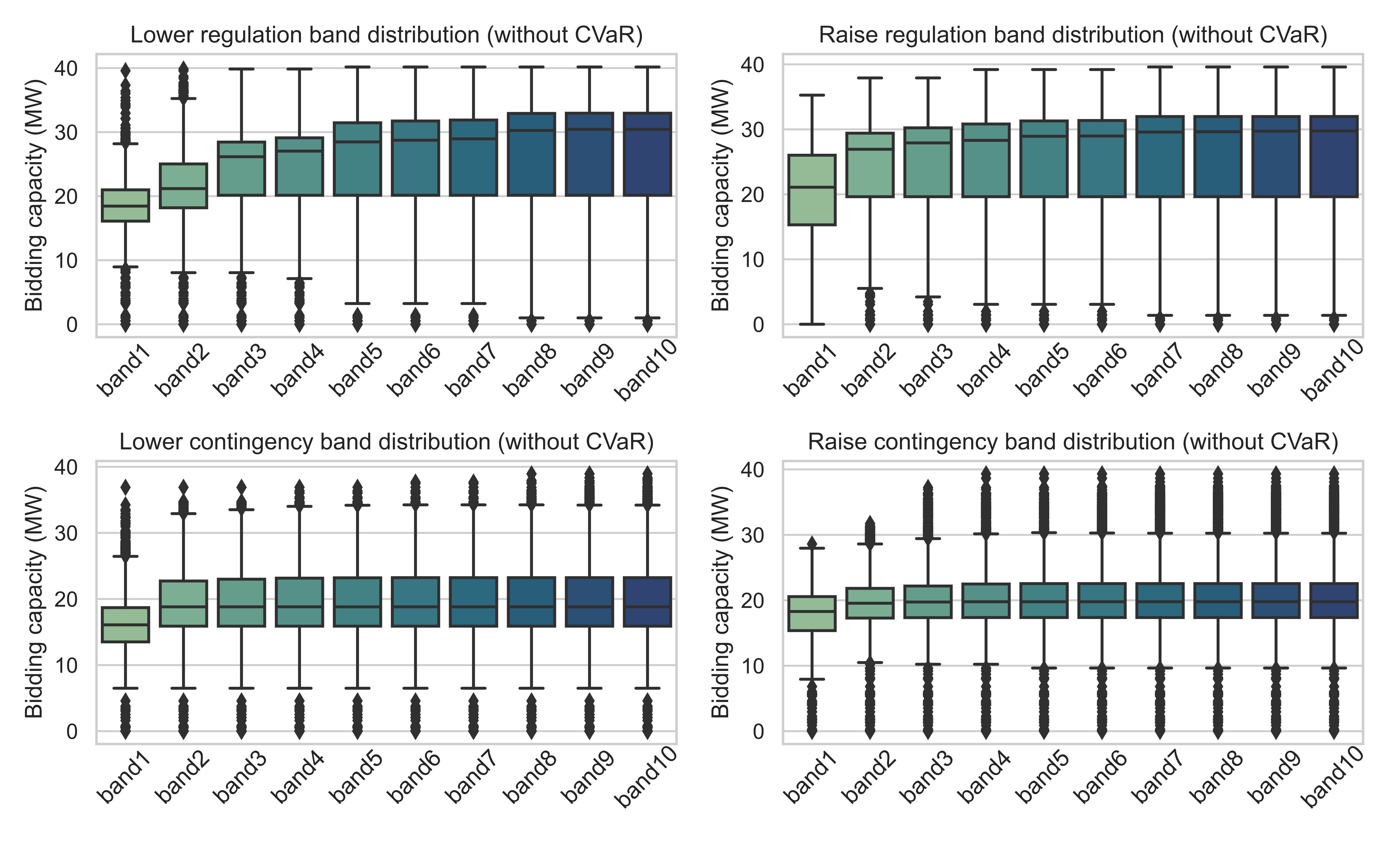}}
\caption{BESS bidding distribution of the original DRL model.}
\label{fig:r6}
\end{figure}

Fig.~\ref{fig:r6} shows the case of the original DRL model. Compared to the proposed algorithm, the original DRL has a wider distribution both in terms of band dimension and capacity dimension. Specifically, the median on regulation capacity ceases to increase at around band 8, with the lower extreme dropping below 10MW and the upper extreme reaching up to 40MW. A similar pattern is observed in the contingency capacity, where the outliers occur significantly. This wider distribution suggests that the original DRL's strategies are more aggressive, often disregarding the risks associated with bidding failure and potential profit loss. In contrast, the CVaR-DRL algorithm adopts a more conservative strategy by integrating the CVaR, which avoids the heavy profit loss in the worst bidding failure scenario. In terms of bidding profits, the CVaR-DRL in the same testing scenarios shows an increase of \$7,879 and \$1,240 in the regulation and contingency markets respectively, reflecting the profit advantages of risk management in bidding strategies.

\vspace{-0.2cm}

\vspace{-0.2cm}

\subsection{Evaluation of LLMs' Strategies}
The offset of FCAS market patterns over time may cause the DRL agent's original strategy to no longer adapt well to new scenarios and lead to a decline in profitability. Thus, LLMs aim to detect the scenario changes in market patterns online for better timeliness, adjust strategies within a safe profit margin for better reliability, and provide explanations to users for better interpretability.

\begin{figure}[htbp]
\centerline{\includegraphics[width=0.85\linewidth]{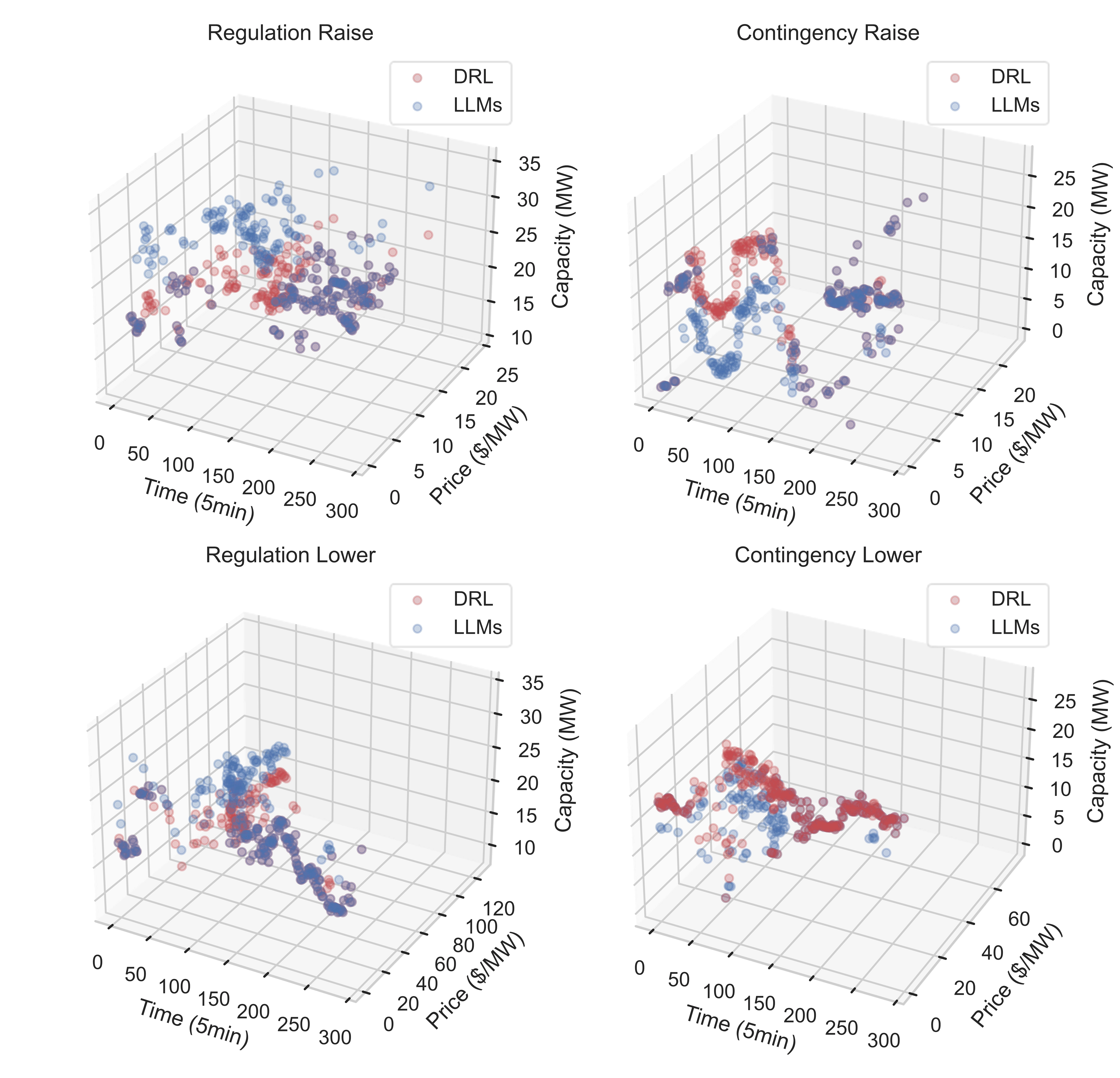}}
\caption{Comparison of LLMs-assisted and DRL original strategies.}
\label{fig:r7}
\end{figure}

Fig.~\ref{fig:r7} compares the bidding results under the LLMs-assisted strategy and the DRL original strategy in the new scenario outside the training set. Typically, in the price axis, the volatility of contingency market clearing prices is significantly higher than that of the regulation market because the demand is often more urgent. However, in this scenario, the regulation market experienced greater volatility. Agent-1 detected the profit opportunities arising from this abnormal market condition and hence in the capacity axis, Agent-2 partially shifted capacity from the contingency markets to regulation markets. In contrast, the original strategy of reserving more contingency capacity based on historical experience would result in profit losses in this scenario, indicating that the LLMs' bidding strategy is more reliable in terms of profitability in such uncommon scenarios. 

In the time axis, the capacity shifting occurred only during the first half of the scenario. For the rest of the period, the regulation prices had returned to common market conditions, thus the original strategy remained essentially unchanged to avoid over-reliance on LLMs' decisions and ensure a safe profit margin considering the possibility of hallucinations. This reflects the ability of LLMs in terms of timeliness to monitor dynamic changes in market conditions online.

\begin{figure}[htbp]
\centerline{\includegraphics[width=0.85\linewidth]{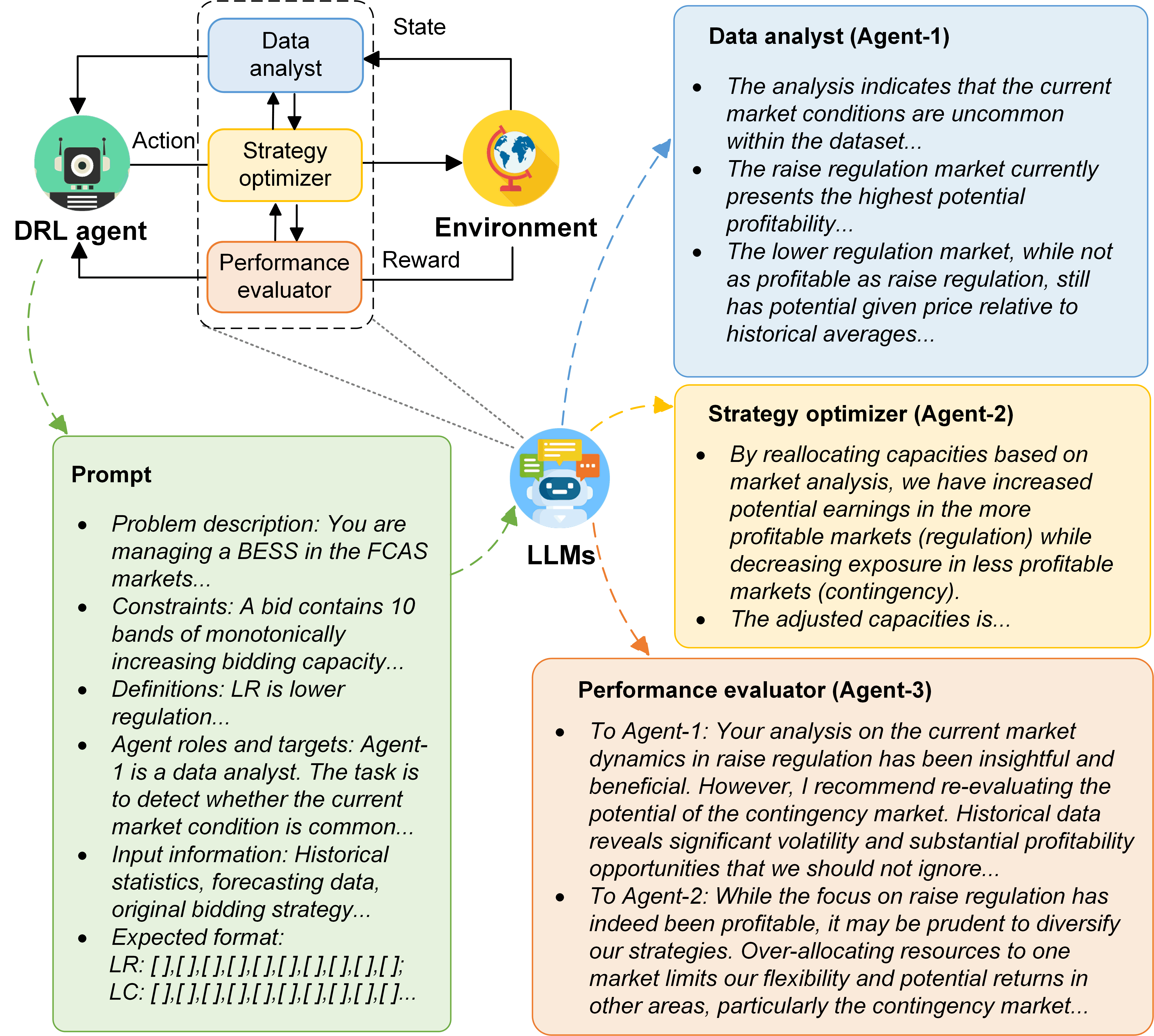}}
\caption{The prompt and answers during interaction with LLMs.}
\label{fig:r8}
\end{figure}

To better demonstrate the workflow of LLMs, we partially excerpt the prompt and answers during interaction with LLMs as shown in Fig.~\ref{fig:r8}. In this process, multiple agents iteratively update their strategies through interactive collaboration based on feedback and affect the interaction between the DRL agent and the environment, thus assisting the DRL agent in making better bidding strategies in new scenarios. In addition, LLMs provide the motivation and rationale for each agent's decision, which enhances the interpretability of the strategies.

\vspace{-0.2cm}

% \vspace{-0.2cm}

\section{Conclusion}
In this paper, we model the complete bidding process of BESS in the joint energy and FCAS markets. We develop the CVaR-DRL algorithm to improve the bidding robustness under aleatoric uncertainties and propose the LLMs-based framework to improve the bidding generalization capabilities under epistemic uncertainties. The experiment results show that the proposed method achieves higher bidding profits compared to the traditional mathematical methods. The CVaR-DRL better reduces the bidding risks caused by forecasting uncertainties. The LLMs enhance the bidding performance of strategies in new scenarios through online market condition analysis, hybrid decision-making, and self-reflection in the ``DRL-LLMs'' loop.

\vspace{-0.2cm}
%\section*{Acknowledgment}
\bibliographystyle{IEEEtran} 
\bibliography{Mybib} 

%%\end{IEEEbiography}

%\end{IEEEbiography}

\end{document}